\documentclass[prd,onecolumn,nopacs,nofootinbib,,preprintnumbers]{revtex4}

\usepackage{mathrsfs}
\usepackage{yfonts}
\usepackage{tipa}
\usepackage{bm}
\usepackage{bbm}
\usepackage{amsmath}
\usepackage{amssymb}
\usepackage{amsthm}
\usepackage{amsfonts}
\usepackage{mathtools}
\usepackage{color}
\usepackage{latexsym}
\usepackage{relsize}
\usepackage[polish,greek,english]{babel}
\usepackage{booktabs}
\usepackage[iso-8859-7]{inputenc}

\usepackage{array}
\newcolumntype{?}{!{\vrule width 2pt}}

\newcommand{\lp}{\left(}
\newcommand{\rp}{\right)}
\newcommand{\lb}{\left[}
\newcommand{\rb}{\right]}

\newcommand{\p}{{\mathrm{P}}}

\newcommand{\lsim}   {\mathrel{\mathop{\kern 0pt \rlap
  {\raise.2ex\hbox{$<$}}}
  \lower.9ex\hbox{\kern-.190em $\sim$}}}
\newcommand{\gsim}   {\mathrel{\mathop{\kern 0pt \rlap
  {\raise.2ex\hbox{$>$}}}
  \lower.9ex\hbox{\kern-.190em $\sim$}}}
\newcommand{\bw}{\begin{widetext}\begin{equation}}
\newcommand{\ew}{\end{equation}\end{widetext}}
\newcommand{\be}{\begin{equation}}
\newcommand{\ee}{\end{equation}}
\newcommand{\ba}{\begin{eqnarray}}
\newcommand{\ea}{\end{eqnarray}}

\newcommand{\diff}{{{\rm d}}}

\newcommand{\bomega}{\boldsymbol{{\omega}}}

\newcommand{\bbe}{\boldsymbol{\rm e}}
\newcommand{\e}{\mathrm{e}}
\newcommand{\ie}{\text{\textschwa}}
\newcommand{\bbie}{\boldsymbol{\textbf{\textschwa}}}

\newcommand{\bT}{\mathbf{T}}

\newcommand{\nn}{\nonumber}

\begin{document}

\title{The spectrum of teleparallel gravity}

\author{Tomi Koivisto}
\affiliation{Nordita, KTH Royal Institute of Technology and Stockholm University, Roslagstullsbacken 23, SE-10691 Stockholm, Sweden}
\affiliation{National Institute of Chemical Physics and Biophysics, R\"avala pst. 10, 10143 Tallinn, Estonia}
\author{Georgios Tsimperis}
\affiliation{Department of Physics, Stockholm University, AlbaNova University Centre, SE-106 91 Stockholm, Sweden}

\preprint{NORDITA-2018-99}
\date{\today}

\begin{abstract}

The observer's frame is the more elementary description of the gravitational field than the metric.
The most general covariant, even-parity quadratic form for the frame field in arbitrary dimension generalises the New General Relativity by nine functions of the d'Alembertian operator.
The degrees of freedom are clarified by a covariant derivation of the propagator. The consistent and viable models can incorporate an ultra-violet completion of the gravity theory, an additional polarisation of the 
gravitational wave, and the dynamics of a magnetic scalar potential.  

\end{abstract}

\maketitle

\section{Introduction}

Gravitational waves, a prediction of General Relativity (GR) that was only recently directly confirmed in the experimental data \cite{Abbott:2016blz}, are most often discussed in terms of fluctuations of the metric field.
However, at a more fundamental level the gravitational field in GR has to be understood as the {\it frame field}.
It has been established long ago that the frame field, in four dimensions a.k.a. the {\it vierbein}, or the {\it tetrad}, is necessary for the consistent
gravitational coupling of the electron \cite{Weyl:1929fm}. Moreover\footnote{The teleparallel theory of the frame field has been considered as the gauge theory of the group of translations \cite{Cho:1975dh,Hayashi:1967se,Aldrovandi:2013wha}. The main difficulties of this interpretation (that the connection is not generated by translations, nor minimally coupled to matter) are resolved in the recently ``purified gravity'' \cite{BeltranJimenez:2017tkd} which is not however considered in this paper, but see \cite{Conroy:2017yln}.}, the frame field formulation of GR facilitates the covariant definition of gravitational energy-momentum complex \cite{moller}.

A frame field is a set of $n$ orthonormal vector fields $\{\bbie_a\}_{a=0,1,\dots, n-1}$
defined on an $n$-dimensional Lorentzian manifold that is interpreted as a model of spacetime. 
All tensorial quantities on the manifold can be expressed using the frame field $\bbie_a$ and its dual coframe field $\bbe^a$. 
In particular, the components of the contravariant metric tensor, $g^{\mu\nu}$, are obtained from the components $\ie_a{}^\mu$ of the frame field $\bbie_a=\ie_a{}^\mu\partial_\mu$ using the Cartan-Killing form $\eta^{ab}$ that is interpreted as the Minkowski metric in the tangent space, as $g^{\mu\nu}=\eta^{ab}\ie_a{}^\mu\ie_b{}^\nu$. The generic frame field has $n^2$ independent components, and it gives rise to the $\frac{1}{2}n(n+1)$ independent components of the symmetric rank-2 tensor $g_{\mu\nu}$. In the case of GR, the remaining $\frac{1}{2}n(n-1)$ components are eliminated by Lorentz invariance. 

It is of interest to consider more general theories that may be formulated in terms of the frame field. 
During the past hundreds years, a plethora of alternatives and extensions to GR have been introduced, and such are currently under extensive investigation especially motivated by the problems of modern 
cosmology \cite{Heisenberg:2018vsk}. One of the first things to check about a new theory of gravity is its inherent consistency and observational viability in the limit of Minkowski space. Our aim is to  classify the possible theories for the frame field in this limit by the properties of the propagator. The poles of the propagator determine the particle content of the theory. 

We shall study the most general frame field action that is Poincar\'e and parity invariant. The main conclusion will be that there exist viable frame field theories which are not captured by the generic metric theory either in Riemannian \cite{Biswas:2011ar} or non-Riemannian \cite{Conroy:2017yln} geometry. 
When restricting to the second order in derivatives, the action reduces to the well-known
M{\o}ller-Pellegrini-Pleba{\'n}ski theory \cite{moller,article} a.k.a. the New GR \cite{Hayashi:1979qx}, whose linearisation has been often considered previously 
\cite{MuellerHoissen:1983vc,Kuhfuss:1986rb,ortin,Karananas:2014pxa,Vasilev:2017twr,Hohmann:2018jso,Hohmann:2018xnb,Blagojevic:2018dpz}. 
The most general action, at the relevant quadratic limit, extends the three-parameter case of New GR by nine functions of the covariant derivative operator, and can accommodate also the ghost- and singularity-free structure that has been previously realised in the metric theories \cite{Biswas:2011ar,Conroy:2017yln}. Besides the spin-2 graviton with the infinite-derivative structure and the spin-0 dilaton-like particle that are expected in the closed string field theory, the spectrum of a consistent frame field theory may also feature the spin-0 Kalb-Ramond-like particle.

In the remainder of this brief paper, we shall report the most general quadratic action, Eq.(\ref{action}), and the field Eqs.(\ref{efes},\ref{efea}) in Sec. \ref{efe},
the propagator, Eq.(\ref{propagator}), in Sec. \ref{propa}, and then present our conclusions in Sec. \ref{appl} and Sec. \ref{conclu}.
 
\section{Field equations} 
\label{efe}
 
The motivation is to uncover the possible properties of gravitation that may not be described by solely the metric. 
We are interested in the most general theory for the (co)frame field $\e^a{}_\mu$, but the properly invariant formulation \cite{Golovnev:2017dox} also includes the 
spin connection $\bomega^a{}_{b}$, though it is purely inertial \cite{Aldrovandi:2013wha,Krssak:2015oua,Golovnev:2017dox}. 
The field strength of the coframe field is written in the differential form notation as $\bT^a = \text{D} \bbe^a = \diff \bbe^a + \bomega^a{}_b\wedge \bbe^b$, but we shall be explicit with the components such as the $T^a{}_{\mu\nu}$ in
\be
\bT^a=\frac{1}{2}T^a{}_{\mu\nu}\diff x^\mu\wedge \diff x^\nu = \lp\partial_{[\mu}\e^a{}_{\nu]} + \omega^a{}_{b[\mu}\e^b{}_{\nu]}\rp\diff x^\mu\wedge\diff x^\nu \,.
\ee
The most general theory that is quadratic in this field strength  
can be parameterised by nine independent functions of the d'Alembertian operator 
$\Box = g^{\mu\nu}\nabla_\mu\nabla_\nu$. We write action for the theory as
\be \label{action}
I = -\int \diff^n x \e L + I_{(M)}\,,
\ee
where $\e=\det{\e^a{}_\mu}=\sqrt{\det{g_{\mu\nu}}}$, $I_{(M)}$ is the action for the matter fields, and the gravitational Lagrangian is
\ba
L & = &  T^{\alpha\mu\nu}\lb c_1(\Box) T_{\alpha\mu\nu} 
   +  c_2(\Box)T_{\nu\mu\alpha} 
   + g_{\alpha\nu} c_3(\Box)T_{\mu}\rb \nn  \\
   & + & T^{\alpha\mu\nu}\Box^{-1}\lb c_4(\Box) \nabla_\alpha\nabla^\beta T_{\beta\mu\nu}
   + c_5(\Box)  \nabla_\alpha\nabla^\beta T_{\mu\nu\beta}
   + c_6(\Box) \nabla_\alpha\nabla^\beta  T_{\mu\nu\beta} 
   + c_7(\Box)\nabla_\nu\nabla^\beta T_{\mu\alpha\beta}
   + g_{\alpha\nu}c_8(\Box)\nabla_\mu\nabla^\beta T_\beta \rb \nn \\
   & + & T^{\alpha\mu\nu}\Box^{-2}c_9(\Box)\nabla_\alpha\nabla_\mu\nabla^\rho\nabla^\sigma T_{\rho\sigma\nu}\,. \label{lagr}
\ea
We have defined the trace $T_\mu=T^\alpha{}_{\mu\alpha}$. 
This action reduces to the three-parameter 
New GR \cite{Hayashi:1979qx} when $c_1(\Box)=c_1$, $c_2(\Box)=c_2$ and $c_3(\Box)=c_3$ are constants, and the rest of the functions
are zero. Assuming the quadratic torsion is modulated by analytic functions, the four terms in the second line are at least fourth derivative and the two terms in the third line are at least sixth order derivative. The pure-gauge connection $\bomega^a{}_{b}$ is given by a Lorentz transformation $\Lambda^a{}_b$ of the Weitzenb\"ock connection ($\bomega^{a}{}_b=0$) as 
$\bomega^a{}_{b} = (\Lambda^{-1}){}^a{}_c\diff\Lambda^c{}_b$, where both $\bomega_{ab}$ and $\Lambda_{ab}$ are antisymmetric. Thus, the action principle (\ref{action}) is understood as $L=L(\bbe^a,\Lambda^a{}_b)$ \cite{Golovnev:2017dox}. We expand the connection as
\be
\omega^a{}_{b\mu} =  \partial_\mu A^a{}_b + \mathcal{O}(A^2)\,, \quad \Lambda^a{}_b \approx \delta^a{}_b - A^a{}_b+ \mathcal{O}(A^2)\,,
\ee
and the coframe field as
\be
\e^a{}_\mu = \delta^a{}_\mu + B^a{}_\mu\,,
\ee
which implies for the inverse 
\be
\ie_a{}^\mu = \delta_a{}^\mu - B^\mu{}_a + \mathcal{O}(B^2)\,, \quad B^\mu{}_a = \delta_b{}^\mu\delta_a{}^\nu B^b{}_\nu\,.
\ee
At the lowest order, the metric perturbation is given by the symmetric part,
\be
g_{\mu\nu} \equiv \eta_{ab}\e^a{}_\mu \e^b{}_\nu =  \eta_{\mu\nu} + h_{\mu\nu} + \mathcal{O}(B^2)\,, \quad h_{\mu\nu} \equiv 2B_{(\mu\nu)}\,, \quad 
B_{\mu\nu} \equiv \delta_{a\mu}B^a{}_\nu\,,
\ee
and for the antisymmetric part we define invariant combination
\be
b_{\mu\nu} \equiv 2\lp B_{[\mu\nu]} - A_{[\mu\nu]} \rp\,,
\ee
in terms of the antisymmetric perturbations of the frame field and of the pure-gauge field. It is then straightforward though tedious to expand the action (\ref{action}) to the second order in the perturbations. Describing the matter action $I_M$ with the linear source term $\tau^{\mu\nu}$, we obtain
\be
I = -\frac{1}{4}\int \diff^n x \lp L_{(h^2)} + L_{(hb)} + L_{(b^2)} \rp + 2\int \diff^n x B_{\mu\nu}\tau^{\mu\nu} + \mathcal{O}(B^3)\,, 
\ee
where the purely metric part is (we denote the trace $h=\eta^{\mu\nu}h_{\mu\nu}$)
\be \label{hh}
L_{(h^2)} = h^{\mu\nu}\lb a(\Box) \Box h_{\mu\nu} + 2b(\Box)\partial^\alpha\partial_\mu h_{\nu\alpha}
+ c(\Box)\lp \partial_\mu \partial_\nu h + \eta_{\mu\nu}\partial_\alpha\partial_\beta h^{\alpha\beta}\rp
+ \eta_{\mu\nu} d(\Box)\Box h + \frac{f(\Box)}{\Box}\partial_\mu\partial_\nu\partial_\alpha\partial_\beta h^{\alpha\beta}\rb\,, 
\ee
there appears the one possible interaction term 
\be
L_{(hb)} = -2h^{\mu\nu} x(\Box)\partial_\mu\partial^\alpha b_{\alpha\nu}\,, 
\ee
and the part involving only the antisymmetric perturbation is
\be
L_{(b^2)} = b^{\mu\nu}\lb y(\Box)\Box b_{\mu\nu} + 2z(\Box)\partial_\mu\partial^\alpha b_{\alpha\nu}\rb\,.
\ee
The functions in (\ref{hh}) read (omitting the arguments of $\Box$ from now on) 
\begin{subequations}
 \label{abc}
\ba
a & = & 2c_1+c_2-c_6-c_7\,, \\ 
b & = & -c_1-\frac{1}{2}\lp c_2- c_3  + c_4 - c_5 - c_6 - 2c_7 + c_9\rp \,, \\
c & = & -c_3+c_8\,, \\
d & = & c_3 - c_8\,, \\
f & = & c_4 - c_5 - c_7 - c_8 + c_9\,.
\ea
\end{subequations}
The rest of the functions are specified as
\be \label{x}
x  =  2c_1 + c_2 + c_3 - c_4 + c_5 - c_6 - c_9\,,
\ee
and
\begin{subequations}
\label{ab}
\ba
y & = & -c_1+\frac{3}{2}c_2+\frac{1}{2}\lp c_3-c_4+c_5+c_6\rp+c_7-\frac{1}{2}c_9\,, \\
z & = & 2c_1-c_2-c_6-c_7\,. 
\ea
\end{subequations}
The field equations for the symmetric part, including a source term, are
\be \label{efes}
-2\tau_{(\mu\nu)} = a \Box h_{\mu\nu} + 2b \partial^\alpha\partial_{(\mu} h_{\nu)\alpha}
+ c\lp \partial_\mu \partial_\nu h + \eta_{\mu\nu}\partial_\alpha\partial_\beta h^{\alpha\beta} 
- \eta_{\mu\nu}\Box h\rp + \frac{f}{\Box}\partial_\mu\partial_\nu\partial_\alpha\partial_\beta h^{\alpha\beta} - x\partial_{(\mu}\partial^\alpha b_{\nu)\alpha}\,,
\ee
and the antisymmetric components of the field equations are
\be \label{efea}
2\tau_{[\mu\nu]} = y \Box b_{\mu\nu} - 2z\partial_{[\mu}\partial^\alpha b_{\nu]\alpha} + x\partial_{[\mu}\partial^\alpha h_{\nu]\alpha} \,.
\ee
The divergence of the symmetric source becomes
\be
-2\eta^{\mu\rho}\partial_\mu\tau_{(\rho\nu)} = (a+b)\Box h^\mu{}_{\nu,\mu} + (b+c+f)h^{\alpha\beta}{}_{,\alpha\beta\nu} - \frac{1}{2}x\Box\partial^\mu b_{\mu\nu} 
=\frac{1}{2}x\lp \Box h^\mu{}_{\nu,\mu} - h^{\alpha\beta}{}_{,\alpha\beta\nu} -  \Box\partial^\mu b_{\mu\nu}\rp \,. 
\ee
In the second equality we have taken into account the relations $a+b=-(b+c+f)=x/2$ that follow identically from the definitions (\ref{abc}) and (\ref{x}). 
Thus, if the coupling of the $h_{\mu\nu}$ and the $b_{\mu\nu}$ vanishes, $x=0$, the usual covariant conservation of energy-momentum is recovered. 
The divergence of the antisymmetric source is
\be
-2\eta^{\mu\rho}\partial_\mu\tau_{[\rho\nu]} = \lp y + z\rp \Box \partial^\mu b_{\mu\nu} 
- \frac{1}{2}x \lp \Box h^\mu{}_{\nu,\mu} - h^{\alpha\beta}{}_{,\alpha\beta\nu}\rp = \frac{1}{2}x\lp  \Box \partial^\mu b_{\mu\nu} - \Box h^\mu{}_{\nu,\mu} + h^{\alpha\beta}{}_{,\alpha\beta\nu}\rp \,.
\ee
In the second equality, we have used that $y+z=x/2$, as dictated by the coefficients (\ref{x}) and (\ref{ab}).
Combining the two divergences shows that, to the linear order in perturbations, we have simply $\partial_\mu\tau^\mu{}_\nu=0$. Another consistency check is that the connection equations of motion are redundant with the 
equations of motion for the antisymmetric frame field perturbation.



\section{Propagator}
\label{propa}

The field $\tilde{B}_{\mu\nu} \equiv B_{\mu\nu}-A_{\mu\nu}$ decomposes into the spin parts
\be
\tilde{B}_{\mu\nu} = (2^+)\oplus(1^+)\oplus(1^-) \oplus(1^-)\oplus(0^+)\oplus(0^+) \equiv (\bf{g})\oplus(\bf{b})\oplus(\bf{m})\oplus(\bf{e})\oplus(\bf{s})\oplus(\bf{w})\,.
\ee
Thus, in terms of the irreducible representations of the rotational group, a rank-2 tensor consists of a tensor piece, one vector and two pseudovector pieces,
and two scalars. Along the lines of Refs \cite{VanNieuwenhuizen:1973fi,MuellerHoissen:1983vc}, they could be referred to as ``gravity'', ``magnetic'', ``momentum'', ``electric'', ``stress'' and ``work'', respectively. 
To construct the spin projection operators \cite{VanNieuwenhuizen:1973fi,MuellerHoissen:1983vc,Kuhfuss:1986rb,Karananas:2014pxa} into the respective subspaces, we define, in terms of the wavevector $k^\mu$, the two bases
\be
\theta_{\mu\nu} = \eta_{\mu\nu} - k_\mu k_\nu/k^2\,, \quad \sigma_{\mu\nu} = k_\mu k_\nu/k^2\,.
\ee
The projection operators we need for the symmetric sector can then be defined as
\ba
\p^{({\bf g})}_{\mu\nu\rho\sigma} & = & \theta_{\mu(\rho}\theta_{\sigma)\nu}-\frac{1}{n-1}\theta_{\mu\nu}\theta_{\rho\sigma}\,, \\
\p^{({\bf m})}_{\mu\nu\rho\sigma} & = & \theta_{\mu(\rho}\sigma_{\sigma)\nu} + \theta_{\nu(\rho}\sigma_{\sigma)\mu} \,, \\
\p^{({\bf s})}_{\mu\nu\rho\sigma} & = & \frac{1}{n-1}\theta_{\mu\nu}\theta_{\rho\sigma}\,. 
\ea
It turns out that we need only one scalar projector, since the $(\bf{w})$-subspace is empty in any possible pure-torsion theory.
When taking into account the antisymmetric sector, the following operators need to be introduced. 
\ba
\p^{({\bf e})}_{\mu\nu\rho\sigma} & = & \theta_{\mu[\rho}\sigma_{\sigma]\nu} - \theta_{\nu[\rho}\sigma_{\sigma]\mu} \,, \\
\p^{({\bf b})}_{\mu\nu\rho\sigma} & = & \theta_{\mu[\rho}\theta_{\sigma]\nu}\,, \\
\p^{({\bf m\times e})}_{\mu\nu\rho\sigma} & = & \theta_{\mu[\rho}\sigma_{\sigma]\nu} + \theta_{\nu[\rho}\sigma_{\sigma]\mu} \,, \\
\p^{({\bf e\times m})}_{\mu\nu\rho\sigma} & = & \theta_{\mu(\rho}\sigma_{\sigma)\nu} - \theta_{\nu(\rho}\sigma_{\sigma)\mu} \,.
\ea
The two first operators form the complete set of orthogonal projectors, and the two last ones mix the symmetric and the antisymmetric
sectors. Then we can rewrite the total field equation using the projections as follows:
\be
k^2\lb a\p^{({\bf g})}_{\mu\nu\rho\sigma} +  (a-3c)\p^{({\bf s})}_{\mu\nu\rho\sigma} + y\p^{({\bf b})}_{\mu\nu\rho\sigma} + \frac{1}{2}x\lp \p^{({\bf e})}_{\mu\nu\rho\sigma} + \p^{({\bf m})}_{\mu\nu\rho\sigma} + \p^{({\bf m\times e})}_{\mu\nu\rho\sigma} + \p^{({\bf e\times m})}_{\mu\nu\rho\sigma}\rp  \rb \tilde{B}^{\rho\sigma} = 2\tau_{\mu\nu}\,.  \label{prop}
\ee
Remarkably, the spin-1 parity-odd electric-momentum subspace $(\bf{m})\oplus(\bf{e})$ has always a degenerate propagator, even when the function $x$ is non-vanishing.
This degeneracy reflects the propagation of an $n$-vector. The symmetry that eliminates it is
\be \label{cartan}
\tilde{B}_{\mu\nu} \rightarrow B_{\mu\nu} + \partial_\mu V_\nu\,,
\ee
whose symmetric part is the diffeomorphism and the antisymmetric part the two-form gauge redundancy. We have one iff we have the other. Interestingly, this $V^\mu$ can be identified as the Cartan's radius vector. It cannot be fully confined into any of the spin subspaces, but it corresponds to some of the components of the ``momentum'' and some of the components of the ``electric'' vector pieces from the symmetric and the antisymmetric sectors, respectively. 
In $n=4$ we may understand that the 4 components of $V_\mu$ are separated into the 2+2 transverse modes of the 2 massless 3-vectors which can only be unleashed in unison.
In fact, in terms of the spin projectors (omitting their indices from now on), we can rewrite $\p^{({\bf m\times e})} + \p^{({\bf e\times m})} = \p^{({\bf m})} + \p^{({\bf e})}$ above, apparently eliminating the degeneracy and the coupling. 
For these non-trivial reasons, we can invert (\ref{prop}) into the propagator $\Pi$ that becomes:
\be
\Pi = \frac{\p^{({\bf g})} }{ak^2}+ \frac{\p^{({\bf s})}}{\lp a-3c\rp k^2} + \frac{\p^{({\bf b})}}{yk^2} + \frac{ \p^{({\bf m})} + \p^{({\bf e})}}{xk^2}\,.  \label{propagator}
\ee  
We have arrived at the main result of this paper. 

\section{Applications} 
\label{appl}

Let us then look at the implications of (\ref{propagator}) in a few different contexts of frame field theories.

\begin{itemize}
\item The {\bf teleparallel equivalent of GR} \cite{Aldrovandi:2013wha,Maluf:2013gaa} corresponds to $c_1=\frac{1}{4}$, $c_2=\frac{1}{2}$, $c_3=-1$, and vanishing higher order terms. These imply $a=c=1$ and $x=y=0$. From the formulas (\ref{abc},\ref{x}) it appears that we may reproduce equivalent theories by many other choices of parameters, but it is important to note that this would require non-analytic functions of the form $c_i \sim 1/\Box$ for $i>3$. Thus, the action of the teleparallel equivalent of GR is unique (up to irrelevant boundary terms) already at the linear order. For further convenience we define this action as $I=\int \diff^n x \e T$, introducing the torsion scalar $T \equiv T_{\alpha\mu\nu}\lp\text{\small{$\frac{1}{4}$}}T^{\alpha\mu\nu} + \text{\small{$\frac{1}{2}$}}T^{\mu\alpha\nu}\rp -T_\mu T^\mu$.

\item The {\bf modified teleparallel $f(T)$ gravity} \cite{Ferraro:2006jd,Linder:2010py} is given by a nonlinear function of the torsion scalar. Such models have received considerable attention in the literature \cite{Cai:2015emx,Hohmann:2017jao}, but nevertheless the nature of their degrees of freedom remains undisclosed, see e.g. \cite{Golovnev:2018wbh,Ferraro:2018axk} for current discussion. There is evidence 
\cite{Ong:2013qja,Chen:2014qtl} that the $f(T)$ models would, in general, contain a propagating extra degree of freedom \cite{Ferraro:2018tpu} or more \cite{Li:2011rn}, but here we confirm the well-known fact that in flat space the propagator reduces to that of GR. That could imply that this class of modified gravity models has a strong coupling problem. Indeed there are disturbing bifurcations in the characteristics \cite{Chen:2014qtl} and constraint structure \cite{Ferraro:2018tpu}.

\item The {\bf modified teleparallel $f(T,B)$ gravity} \cite{Bahamonde:2015zma,Bahamonde:2016grb}, where we have the boundary term $B=\mathcal{D}_\mu T^\mu$ in terms of the metric Levi-Civita connection $\mathcal{D}_\mu$, has been motivated by the relation of the metric Ricci curvature $\mathcal{R}$ and the torsion invariants, $\mathcal{R}=-T+2B$, due to which these models can be also considered as $f(T,\mathcal{R})$ gravity
\cite{Paliathanasis:2017efk,Karpathopoulos:2017arc}. Indeed, we obtain the four functions as $a=f_T$, $c=f_T-f_{BB}\Box$, $x=y=0$, implying that they propagate an extra scalar degree of freedom, in complete analogy to the well-studied $f(\mathcal{R})$ models. The scalar field has the mass $\sim 1/\sqrt{f_{BB}}$, and therefore one should have $f_{BB}>0$ to avoid a tachyonic instability. 

\item The {\bf New GR} \cite{moller,article,Hayashi:1979qx} was considered at the linear order in e.g. Sec. 4.6 of \cite{ortin}, and its field content can be deduced from detailed analyses in the more general context of the Poincar\'e gauge theory \cite{Kuhfuss:1986rb,Karananas:2014pxa,Blagojevic:2018dpz}. The one-parameter class of theories $2c_1+c_2+c_3=0$ i.e. $x=0$ involves $(n^2-3n)/2$ components of $b_{\mu\nu}$ due to the symmetry $b_{\mu\nu}\rightarrow b_{\mu\nu}+\partial_{[\mu}v_{\nu]}$, where $v_\mu$ is an arbitrary vector. However, note that though originating from the ``magnetic'' pseudovector, the Kalb-Ramond field has helicity 0 since at the massless limit, oppositely to the Maxwell field, it is the longitudinal mode that remains whilst the transverse modes decouple.   

\item The generic theory (\ref{lagr}) 
is a {\bf higher-derivative New GR}. 
One should set $a=1$ to obtain the canonical normalisation for the graviton. Now the gravitational wave also may possess a breathing mode, which propagates healthily, given that either $c>1$ or $c<1/3$. 
It is possible to give a mass to this scalar, but not to the graviton nor the scalar particle associated with the Kalb-Ramond field, without introducing ghosts or non-analytic functions $c_i$. 
Again, the crucial symmetry (\ref{cartan}) requires $x=0$, and the Kalb-Ramond field is not a ghost given that $y>0$. The phenomenological viability of these models might be worth investigations.

\item The prototype {\bf infinite-derivative gravity}  \cite{Tomboulis:1997gg,Biswas:2011ar,Modesto:2011kw} is given by $a=c=e^{\frac{-\Box}{M^2}}$, $x=y=0$, where $M^2$ is the energy scale of non-locality. Such gravity theories could avoid both ghosts and singularities, and indeed they are often studied both at classical \cite{Buoninfante:2018xiw} and quantum \cite{Modesto:2017sdr} levels. We note that the teleparallel prototype theory can be realised simply as $I=\int \diff^n x \e e^{\frac{-\Box}{M^2}}T$, whereas in the purely metric formulation the action requires the superposition of the Einstein-Hilbert and a more complicated term that is quadratic in the Riemann curvature \cite{Biswas:2011ar}. 
\end{itemize}

To consider even more general frame field theories, it might be interesting to relax our main assumptions of 1) metric-compatibility 2) parity-invariance or 3) analyticity. To proceed towards nonlinear orders, a natural first step would be to repeat the computation in an (a)dS background. Of course, one can also add further fields besides the frame field. We will present one interesting example, wherein we add a scalar field $\phi$ for the purpose of promoting the previous example into a scale-invariant theory. 
\begin{itemize}
\item An example of a {\bf scale-invariant teleparallel theory} is given (in $n=4$ for simplicity) by
\be
L = \phi^2 \lp \text{\small{$\frac{1}{4}$}}T^{\alpha\mu\nu} + \text{\small{$\frac{1}{2}$}}T^{\mu\alpha\nu} - \text{\small{$\frac{1}{3}$}} g^{\alpha\nu}T^\mu\rp e^{-\phi^2\Box}T_{\alpha\mu\nu} - 6 \lp D_\mu \phi\rp\lp D^\mu \phi\rp\,,
\ee
where the covariant derivative involves the torsion $D_\mu = \partial_\mu - \frac{1}{3}T_\mu$ \cite{Maluf:2011kf,Lucat:2016eze}. This action is invariant under the conformal transformation of the coframe $\e^a{}_\mu \rightarrow e^{\theta}\e^a{}_\mu$ accompanied by the rescaling $\phi \rightarrow e^{\theta}\phi$. We can choose the gauge $\phi=1/M$ in order to explicitly recover the previous case. 
It is not possible to adjust the coefficients above without either breaking the scale-invariance or the symmetry (\ref{cartan}).
\end{itemize}

\section{Conclusion}
\label{conclu}

To summarise our derivations, we deduced that the most general quadratic torsion action contains nine free functions, and found that four of them are independent at the linear order and appear in the propagator (\ref{propagator}).
The purely metric sector of the theory is determined by the two independent functions $a$ and $c$ which describe the propagation of the graviton and the dilaton. Now there is also the function $y$ which determines the propagation of the Kalb-Ramond field, and the function $x$ which controls the non-conservation of matter energy-momentum and the propagation of the Cartan radius vector. 

An issue we didn't touch upon in this paper was raised recently in Ref. \cite{Golovnev:2018wbh}. The degrees of freedom depend upon the background geometry. Which is now the geometry that defines 
a physical observer? We have assumed here that fluctuations around the vacuum $\bbie_a = \boldsymbol{\delta}_a$ occur in the Weitzenb\"ock geometry $\bomega^a{}_b=0$. It appears to be consistent as well to consider that observations take place, say, in the properly parallelised geometry $\bomega^a{}_b= \diff\boldsymbol{\delta}_b(\boldsymbol{B}^a)$. The results in this prescription would be obtained from the above simply by erasing the antisymmetric fluctuations.

 

\acknowledgments{A comment from Jose Beltr\'an helped to correct a mistake about $x\neq 0$ and to considerably improve the paper.}

\bibliography{tetrad}

\end{document}